\documentstyle[preprint,aps,epsf]{revtex}
\tightenlines
\draft
\begin{document}
\title{The dynamics of the spherical p-spin model: 
from microscopic to asymptotic}
\author{Bongsoo Kim \footnote{permanent address: Dept. of
    Physics, Changwon National University, Changwon, 641-773, Korea.}
  \and {Arnulf Latz} \footnote{corresponding author}}
\address{
Institut f\"{u}r Physik, Johannes Gutenberg 
Universit\"{a}t Mainz, Staudinger Weg 7, D-55099 Mainz, Germany}

\maketitle

\begin{abstract}
We have numerically investigated the mean-field dynamics of the 
the $p$-spin interaction spin glass model with p=3 using an efficient 
method of integrating the dynamic equations.
We find a new time scale associated with
the onset of the breakdown of the fluctuation-dissipation theorem in the 
intermediate time regime.
We also find that the off-equilibrium relaxation
exhibits a sub-aging behavior in the intermediate times and crosses over
to a simple aging in the asymptotic regime.  
\end{abstract}
\pacs{75.10.Nr, 61.20.Lc, 64.70.pf}

Since the recognition \cite{kt} of possible deep connection between 
structural glass and some class of spin glass systems such as 
the $p$-spin  model \cite{derr,gm}, 
the spherical $p$-spin model \cite{cs1}, due to its analytic accessibility, 
has been the subject of intense research in both statics and dynamics.
Here we  examine some open questions (such as scaling behavior) 
\cite{bckm1}  of the 
off-equilibrium dynamics of the mean field spherical $p$-spin model
by developing a new method of integrating the dynamic equations of the model.

We consider the system of $N$  spins $S_1, S_2, \cdots S_N$
which interact via Hamiltonian
\begin{equation}
H=- \sum_{1 \leq i_1 < i_2 \cdots < i_p \leq N} J_{i_1i_2\cdots i_p} 
S_{i_1} S_{i_2} \cdots S_{i_p}
\label{eq:1}
\end{equation}
where the spins are continuous variables subject to the spherical constraint
$\sum_{i=1}^{N} S^2_i =N$.
The coupling $J_{i_1i_2\cdots i_p}$ is a Gaussian random variable 
with zero mean and variance
$[J^2_{i_1i_2\cdots i_p}]_J = p!/2 N^{p-1}$.
It is well known \cite{cs1} that this model (with $p>2$) exhibits 
an equilibrium  phase transition at a finite temperature $T=T_S$ 
from paramagnetic phase to spin glass phase characterized by 
one-step replica-symmetry breaking. This temperature is lower than the
than the dynamical freezing temperature $T_D$, below which the dynamics
is non ergodic. 

For the time evolution of the system, we consider the following 
dissipative Langevin dynamics
\begin{equation}
\Gamma^{-1}\partial_t S_i (t) = - z(t)S_i(t)-
\frac{\partial H}{\partial S_i(t)}+\eta_i(t)
\label{eq:2}
\end{equation}
where $z(t)$ is a Lagrange multiplier to enforce the spherical
constraint for all times.
In order to satisfy the detailed balance, the thermal noise $\eta_i(t)$ 
is chosen to be Gaussian with zero mean and variance 
$\langle \eta_i(t)\eta_j(t')\rangle =2\Gamma^{-1} T \delta_{ij}\delta(t-t')$.
The inverse of the kinetic coefficient $\Gamma$ sets  
the microscopic time scale, which henceforth is set to unity.

This Langevin dynamics governs the time evolution of the system
starting from, for example, the random disordered spin configuration 
toward thermal equilibrium. Two physical quantities of interest, 
which quantitatively characterize this time evolution, are 
the two-time correlation function 
$C(t,t_w)=\sum_{i=1}^{N}[<S_i(t)S_i(t_w)>]_J/N$ 
where $\langle \cdots \rangle$ and $[\cdots ]_J$ represent  averages 
over thermal noise and random coupling, respectively,
and the response function 
$R(t,t_w)=\sum_{i=1}^{N}[\frac{\partial <S_i(t)>}
{\partial h_i(t_w)}]_J/N$
where $h_i(t_w)$ is the external field turned on at time $t_w$.
In equilibrium, these quantities
become time-translation invariant, i.e., 
$C(t,t_w)=C(\tau)$ and $R(t,t_w)=R(\tau)$, $\tau \equiv t-t_w$.
Moreover, there exists a fundamental relationship between  the 
correlation and the response, known as the fluctuation-dissipation 
theorem (FDT), which takes the form for the present irreversible dynamics
\cite{dh}
\begin{equation}
R(\tau)=-\frac{\theta(\tau)}{T}\frac{\partial C(\tau)}{\partial \tau}
\label{eq:3}
\end{equation} 
where $\theta(\tau)$ is the unit step function which reflects the
causality for the response function. 
We will see below that the non-equilibrium dynamics of the 
present system manifests an explicit strong dependence of the two times 
$t$ and $t_w$ (aging) and an interesting modification of FDT \cite{h1,cs1,ck}.

At this stage, we find it very useful to introduce an integrated 
response function $F(t,t_w)$ defined as 
\begin{equation}
F(t,t_w)\equiv -\int_{t_w}^t ds \, R(t,s).
\label{eq:4}
\end{equation}
From (\ref{eq:4}) $F(t,t^{-})=0$  and 
$\partial F(t,t_w)/\partial t_w = R(t,t_w)$.
FDT (\ref{eq:3}) then takes the form $F(\tau)=\theta(\tau)[C(\tau)-1]/T$.

In the limit of $N$ going to infinity, one can treat the dynamics 
exactly using the standard functional method such as Martin-Siggia-Rose
formalism \cite{MSR,kt}. 
In particular, the dynamics is governed by the closed
set of coupled equations of $C$ and $F$:
\begin{eqnarray}
 \partial_t C(t,t_w) &=& -z(t) C(t,t_w)+\frac{p(p-1)}{2} 
\int_0^t ds C^{p-2}(t,s) (\partial_s F(t,s)) C(t_w,s) \nonumber \\
                     &+& \frac{p}{2}\int_0^{t_w}ds C^{p-1}(t,s)
\partial_s F(t_w,s),
\label{eq:5}
\end{eqnarray}
\begin{equation}
 \partial_t F(t,t_w)= -1 -z(t) F(t,t_w)+\frac{p(p-1)}{2} \int_{t_w}^t ds 
C^{p-2}(t,s)(\partial_s F(t,s)) F(s,t_w).
\label{eq:6}
\end{equation}
where $z(t)\equiv T-pE(t)$, $E(t)$ being the average 
energy per spin.  The energy density $E(t)$ is related to  $C$ and $F$ as
$E(t)=-(p/2) \int_0^t ds \, C^{p-1}(t,s)\partial_s F(t,s)$. 
The same type of the equations  have been
derived in various physical contexts
 such as the dynamics of a long-range superconducting wire 
network ($p=4$)\cite{cfik}, 
the dynamics of Amit-Roginsky model($p=3$) \cite{fh}, and 
the dynamics of a particle in the random potential in large 
dimensional space \cite{h2,fm,cl,h3}.

As Cugliandolo and Kurchan (CK) \cite{ck} have shown, 
the dynamics exhibits the two distinct regimes depending on the 
relative magnitude of the two times $\tau (t \equiv \tau+t_w)$ and $t_w$.
The first regime  (while $t,t_w \rightarrow \infty$, the time difference
$\tau \equiv t-t_w$ is finite, i.e., $\tau/t_w \rightarrow 0$)
is the regime where the time-translation invariance and FDT hold.
That is,  $C(\tau+t_w,t_w)=C(\tau)$, $F(\tau+t_w,t_w)=F(\tau)$ and 
$T F(\tau)=(C(\tau)-1)$ independent of a given $t_w$. 
In this regime, the dynamic equation for $C(\tau)$ can be easily derived 
from the equation for the integrated response and is given by
\begin{equation}
(\partial_{\tau}+T) C(\tau)
+p\left( E_{\infty}+\frac{1}{2T} \right)(1-C(\tau))+\frac{p}{2T}
\int_0^{\tau} ds C^{p-1}(\tau-s)\frac{dC(s)}{ds}=0.
\label{eq:7}
\end{equation}
where $E_{\infty}$ is the long time limit of the energy density $E(t)$,
$E_{\infty} \equiv \lim_{t\rightarrow \infty}E(t)$.
One can recognize that this equation is quite similar to
a schematic model developed in the context of the mode-coupling 
theory (MCT) for the glass transition in structural glasses \cite{gl}.
In particular, apart from the term involving $(1-C(\tau))$, 
(\ref{eq:7}) with $p=3$ is identical
to the Leutheusser model (without the inertial term) \cite{leuth}.
Hence one can sense that this equation, as in the Leutheusser model, 
may lead to a dynamic transition to a nonergodic phase where
the long-time limit of $C(\tau)$ is non-vanishing. 
It is easy to show from (\ref{eq:7}) that
the non-ergodicity parameter  $q\equiv \lim_{\tau \rightarrow \infty}C(\tau)$ 
is then related to $E_{\infty}$ via 
\begin{equation}
-(T-pE_{\infty})+\frac{p}{2T}(1-q^{p-1})=-\frac{T}{1-q}
\label{eq:8}
\end{equation}
But the full understanding of the FDT dynamics requires 
the asymptotic value of the energy density, $E_{\infty}$, for which
one  has to consider the dynamics in different regime (aging regime);
the two regimes are closely coupled to each other.

When the two times $t$ and $t_w$ are large and well separated, 
the relaxation is very slow, and hence the time derivatives
in (\ref{eq:5}) and (\ref{eq:6}) can be ignored.
In this situation, it is found that 
the scaling Ansatz for $C$ and $F$
\begin{equation}
C(t,t_w)=  {\cal C}\left[\frac{h(t_w)}{h(t)}\right], \,\,\,
F(t,t_w)= {\cal F}\left[\frac{h(t_w)}{h(t)}\right] 
\label{eq:9}
\end{equation}
leads to
\begin{equation}
\frac{p(p-1)}{2T^2}(1-q)^2q^{p-2}=1,
\label{eq:10}
\end{equation}
\begin{equation}
T {\cal F}(\lambda)=x{\cal C}(\lambda)-[1-(1-x)q],
\label{eq:11}
\end{equation}
and 
\begin{equation}
E_{\infty} = -\frac{1}{2T}\left[ 1-(1-x)q^p \right]
\label{eq:12}
\end{equation}
with $x=(p-2)(1-q)/q$ for non vanishing $q$.
The equation (\ref{eq:11}) with $x <1$  is a 
 modification of FDT in aging regime and
 is one of the most important results obtained from the asymptotic analysis.
Note that although the actual dynamics will select the form of $h(t)$
uniquely, the above asymptotic analysis holds for an arbitrary monotonically
increasing function $h(t)$ (known as the time-reparametrization invariance):
the function $h(t)$ remains undetermined within 
the asymptotic analysis.  

We  now go back to the equation (\ref{eq:7}) and discuss the 
dynamics of FDT regime.
First note from (\ref{eq:8}) that $q=0$ is always a solution for all
temperatures.  We see from (\ref{eq:10}) that  the non-vanishing $q$ starts to 
appear at the temperature  $T^{\ast}=[2p^{1-p}(p-1)(p-2)^{p-2}]^{1/2}$
at which $q=q^{\ast}=(p-2)/p$ ($q^{\ast}=1/3$ and $T^{\ast}=2/3$ for $p=3$).
Thus below $T^{\ast}$,  starting from the initial state 
with {\em zero} energy density \cite{bbm} ({\em e.g.}, 
the state with all spins up), the system always chooses 
the solution with {\em higher} energy
(which is the highest TAP state at a given $T$ \cite{ck,kpv,cs2,fp}) 
among these two solutions.
Thus for $T_D < T \leq T^{\ast}$, $q=0$ is the genuine solution, and
$E_{\infty}=-1/(2T)$, $x=1$. The system is ergodic.
The dynamic transition temperature $T_D$ is determined by
$x(T=T_D)=1$ for $q\neq 0$. This condition with
(\ref{eq:10}) lead to $q_D = (p-2)/(p-1)$ and 
$T_D =[p(p-2)^{p-2}/(2(p-1)^{p-1})]^{1/2}$
(for $p=3$, $T_D=\sqrt{3/8} \simeq 0.612 \cdots$ and $q_D=1/2$).
Therefore the term involving $(1-C(\tau))$ drops
out in (\ref{eq:7}) and the resulting equation 
is the same as a schematic mode-coupling equation for supercooled
liquids. Therefore the relaxation dynamics above the dynamic transition  
exhibits the well-known scaling laws derived with MCT \cite{goetze}.
For $T \leq T_D$ the system chooses $q \neq 0$, $x < 1$, and 
$E_{\infty}+1/(2T) >0$. Hence the system is non-ergodic.
Thus the term $p(E_{\infty}+1/2T)(1-C(\tau))$ in (\ref{eq:7}) is 
turned on. 
This additional term makes the dynamics in the FDT regime  
in the present model differ from that of the MCT for supercooled liquids.
In particular, near the transition $q(T)$ shows a linear behavior
$q(T)=q_D + \mbox{const.}(T_D-T)$ instead of a square-root singularity
observed in MCT. Also whereas the critical relaxation 
is seen very near and at the transition
in MCT, in the present situation the correlation exhibits
a critical relaxation $C(\tau)=q+\mbox{const.}\tau^{-a}$ for
{\em all} temperatures below the transition 
with the exponent $a$ related to the FDT violating
factor $x$ as $\Gamma^2[1-a]/\Gamma[1-2a]=x/2$ \cite{cl}, $\Gamma$
being the gamma function.

Though the asymptotic analysis put forward by CK 
provided quantitative informations on the both FDT and the
asymptotic  aging dynamics, it leaves
unexplored the non-equilibrium dynamics in the intermediate time regime.
Moreover, the time reparametrization invariance 
leaves undetermined the form of the function $h(t)$ which 
the original set of causal dynamic equations would select in the course of
its evolution. Therefore, we do not know what type of scaling feature the
long time aging dynamics might exhibit. 

In this work, we have developed an 
efficient way of integrating the mean-field dynamic equations
({\ref{eq:5}) and (\ref{eq:6}), which
can extend the solution to the time regime long enough so that one can  
clearly investigate the aforementioned important open questions.
Basic idea is that since the relaxation becomes very slow at long times
one can employ an adaptive integration time step \cite{fghl}.
There are two crucial ingredients in the present method.
First, one has to work with the integrated response  instead
of the response function since  the integrated response
relax much more slowly than the response (which is basically 
a time derivative of the correlation).
Another ingredient is to increase the integration time step along the both
two time directions in such a way that the fast relaxing time regime 
should have many integration points. 
The details of the present method will be given elsewhere.

Now we turn to the discussion of the numerical results.
All presented results are for $p=3$.
Here we focus on the dynamics below the transition. 
Shown in Fig.~1(a) is  $C$ and $F$ for $T=0.5$. 
Both functions exhibit a strong $t$-dependence (aging effect)
which persists up to longest time $t$. 
In contrast to the situation above the transition, here the aging will 
not be interrupted, and the system remains perennially out-of-equilibrium.
The parametric plot $-TF(t,t-\tau)$ versus $C(t,t-\tau)$ for 
$T=0.5$ shown in Fig.~2(b)  exhibits
several interesting features. After some time $t \sim 10^2$, independent of
time $t$,  the FDT is established in the range $q \leq C \leq 1$.
Then, after still longer time $t \sim 10^5$,  
all the curves with different times $t$  merge into a single straight
line  with slope close to the FDT violation parameter $x \simeq 0.565 \cdots$.
Note the huge time interval (more than 9 decades!) it takes 
 to reach the asymptotic regime.
Therefore in the asymptotic regime, the FDT and its violation can be
characterized by the two straight lines with slopes $1$ and $x$
meeting at $C=q$: $-TF=1-C$ for $q \leq C \leq 1$ and 
$-TF=x(1-C)+(1-x)(1-q)$ for $C\leq q$. This is 
in accordance with the result of the asymptotic analysis (\ref{eq:11}).
Similar breakdown of FDT has been observed in simulations of 
supercooled liquids \cite{parisi,kb,lapr} and spin glass \cite{rf}.

One interesting new feature we find in the off-equilibrium dynamics 
in the intermediate time region is the existence of a new
time scale associated with the onset of FDT violation.
Note that  the breakdown of FDT occurs sharply at $C=q$ (see Fig.~2(b)).
Hence we are able to measure  this time scale by reading off the 
times $\tau_p$ defined as $C(t,t-\tau_p)=q$ for each fixed $t$.
The double-log plot $\tau_p$ versus $t$ in the inset of Fig.~2(a)
indeed demonstrates that $\tau_p$ shows a power law dependence on $t$
as $\tau_p \sim t^{\phi}$ with $\phi \simeq 0.68$. 
Furthermore, the big difference in $\tau_p$ and $t$ for large $t$
implies that $\tau_p \sim t_w^{\phi}$ with the same exponent.
Since $\phi$ is smaller than $1$, we find  a new time scale 
characterizing the aging dynamics in intermediate time region.
It is thus highly desirable to have a theoretical development
which can provide a detailed information on the dynamics in the 
intermediate times and on the crossover to the asymptotic time regime.
The presence of this time scale in the spherical SK model ($p=2$)
was demonstrated by Zippold et al \cite{zkh}.

We now discuss the scaling behavior in the aging regime. 
Recall that  the asymptotic analysis
loses information on the unique selection of $h(t)$.
The scaling property in our numerical solution is important 
since  it will be able to determine the specific form 
of the function $h(t)$ that the system actually selects.
Figure 2(a) shows the relaxation of the correlation function 
$C(\tau+t_w, t_w)$ for different waiting times $t_w$ at $T=0.5$.
Due to the adaptive integration steps along the both $t$ and $t_w$ axes, 
it is inevitable that the short time data  for each fixed $t_w$ is lost.
It is natural to first examine the possibility of $h(t)$ being a power of $t$, 
$h(t) \sim t^{\gamma}$. This means that  
the correlation in the aging regime shows a scaling behavior  
$C_A(\tau+t_w,t_w)={\hat C}(\tau/t_w)$ (known as the simple aging).
To check this scenario, we plot in Fig.~2(b) the correlation function 
$C(\tau+t_w,t_w)$ against the rescaled time $\tau/t_w$. 
Indeed, we observe that the relaxation data with longest waiting times
collapse onto a single scaling curve. Thus we find that the relaxation
obeys  the simple aging in the asymptotic time regime.
It is also very interesting to observe that the relaxation in the 
intermediate waiting times exhibit a sub-aging behavior (the characteristic 
relaxation time grows slower than the waiting time),
which is often observed \cite{vhob}in the experimental data of thermoremanent
magnetization (TRM) of real spin glass systems.
Note that the crossover from the sub-aging to the simple aging
occurs over a rather broad time region.
We find that close to the dynamic transition, as shown in Fig.~3(a), 
the simple scaling is not yet fulfilled for waiting times 
which are long enough to reach simple aging at $T=0.5$. 
This suggests that the crossover time from sub-aging to simple 
aging becomes broader as the transition is approached. 

Though the sub-aging eventually makes a crossover to the simple 
aging behavior, due to the broadness of the crossover regime 
(particularly near the transition), 
we wanted to know whether there is a scaling function $h(t)$, which  
can collapse the data in the intermediate times.
One available empirical form is 
$h(t)=\exp[\frac{1}{1-\mu}(t/t_0)^{1-\mu}]$,
which has been successfully used to collapse the TRM data
in spin glasses \cite{vhob}.  
The time $t_0$ is the microscopic time scale ($t_0=1$ for our case).
The case of $\mu=0$ corresponds to the absence of aging
(no waiting time dependence), and the case of $\mu=1$ to the
simple aging. Thus, the sub-aging behavior will give $0<\mu<1$. 
As shown in Fig.~3(b), with this empirical form of $h(t)$ with
$\mu \simeq 0.81$ for $T=0.61$, we can collapse {\em both} 
the correlation and the integrated response with largest waiting times. 
We find that the exponent $\mu$ becomes larger in order to collapse
the sub-aging data (with the same range of waiting times) 
at lower temperatures.  
We should emphasize that $\mu$ is an {\em effective} exponent, i.e., 
it tends to {\em increase} in order to incorporate the data with
longer waiting times into scaling collapse due to the crossover to
simple aging.

In summary, we have developed an efficient method of 
integrating the dynamic equations of a mean field spin glass model 
with $p$-spin interaction. 
This method allows us to see numerically for the first time 
the entire dynamic regime covering from microscopic to asymptotic regime. 
Tio cover such a broad dynamic range will also be of major importance
in the sudy of aging in structural glasses \cite{latz00}.
We observe several new dynamic features in the
intermediate time regime including the existence of a new time
scale at which the breakdown of FDT sets in, the broad sub-aging
regime and its crossover dynamics to the simple aging.
These findings offer a new theoretical challenge for
further understanding of the off-equilibrium dynamics of the $p$-spin
model.

\acknowledgments
We thank K. Binder, L. Cugliandolo, H. Horner, K. Kawasaki, W. Kob, J. Kurchan,
M. M\'{e}zard, R. Schilling and  T. Theenhaus for valuable discussions and 
suggestions. B.K. thanks R. Schilling and K. Binder for hospitality
during his sabbatical at Mainz.
This work was supported by the SFB 262 and the MWFZ of the Johannes
Gutenberg-University of Mainz.

\newpage

\begin{figure}
\epsffile{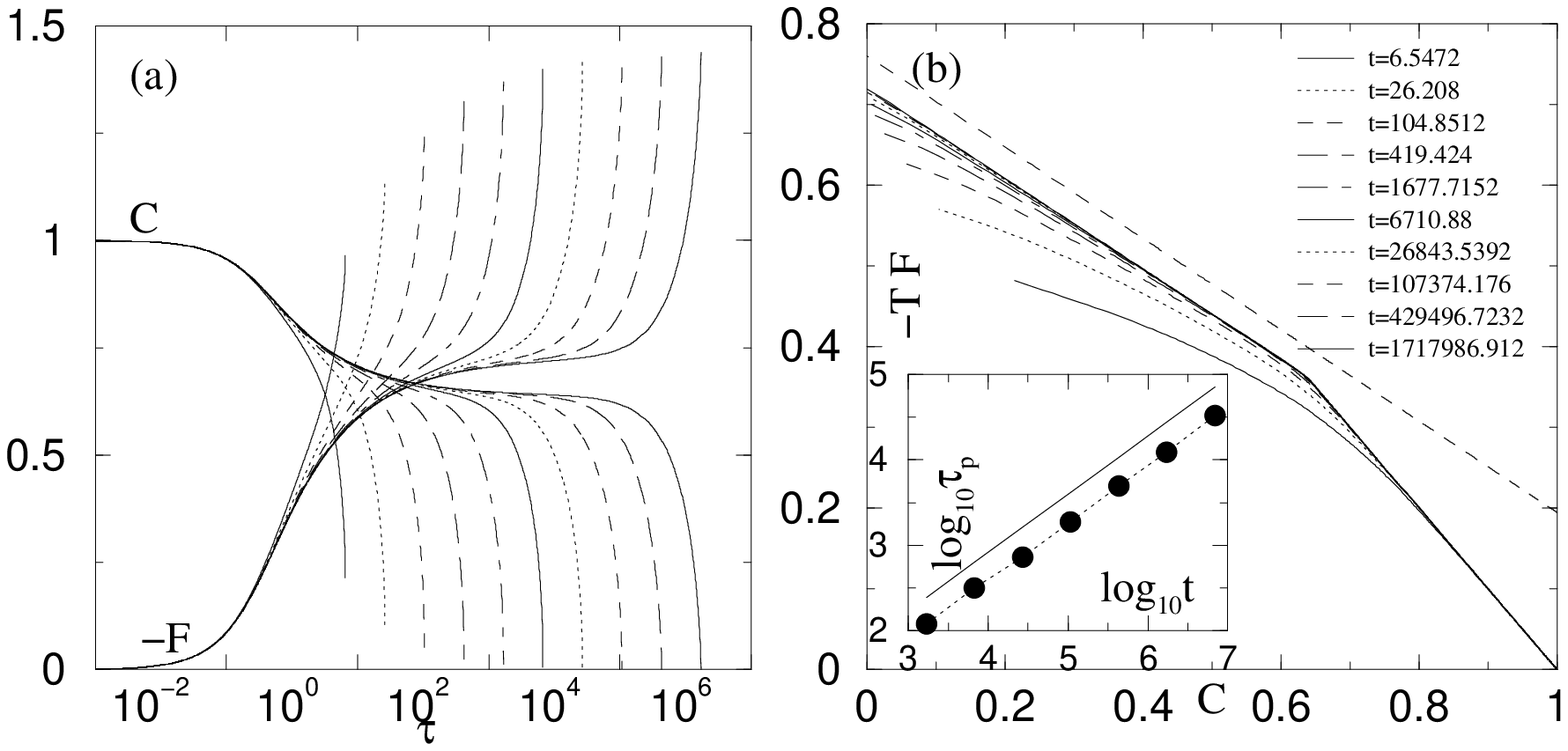}
\caption{(a) The correlation function $C(t,t-\tau)$
and the integrated response function $F(t,t-\tau)$ versus $\tau$ with
different $t$ for $T=0.5$. (b)The parametric plot $-TF(t,t-\tau)$
versus $C(t,t-\tau)$ with different $t$ as in (a). The dashed 
straight line with slope $-0.565$ is shown for comparison.
Inset: $\tau_p$ versus $t$ in a log-log plot.
The solid straight line with slope $0.68$ is shown for comparison.}
\label{f.1}
\end{figure}

\begin{figure}
\epsffile{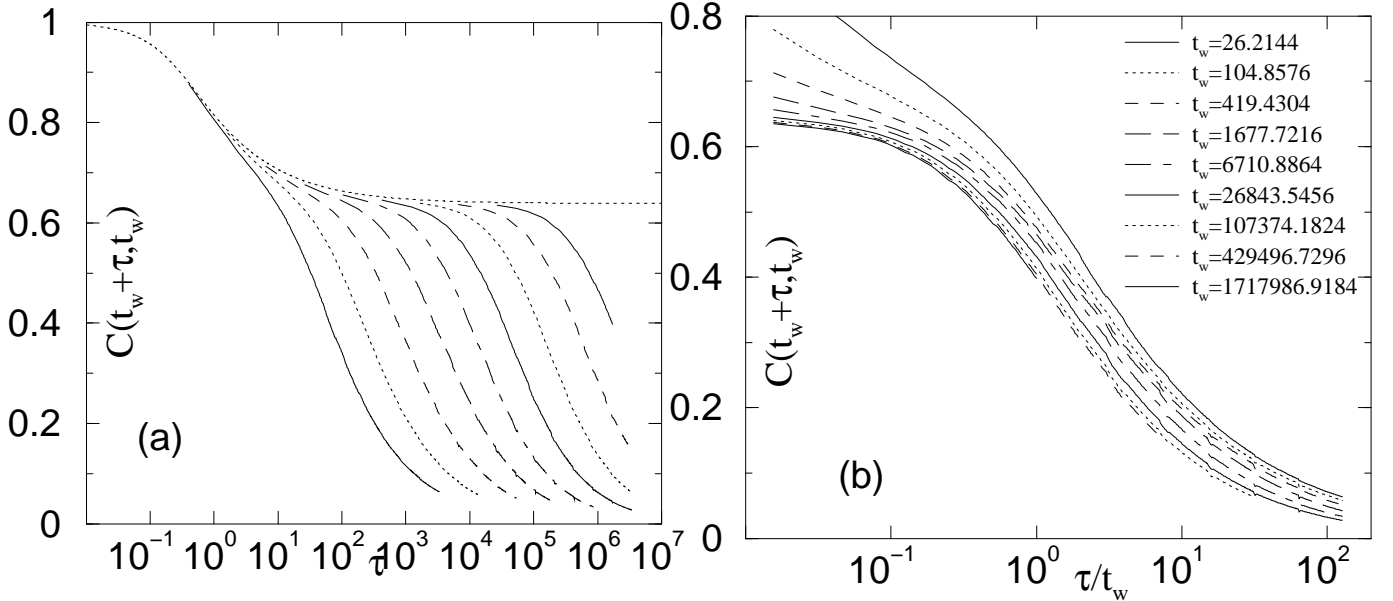}
\caption{(a) The correlation function $C(t_w+\tau, t_w)$ versus
$\tau$ for different waiting times $t_w$ at $T=0.5$. 
The upper dotted line is the numerical solution of (7).
(b) $C(t_w+\tau,t_w)$ versus rescaled time $\tau/t_w$ for the data
shown in (a).}  
\label{f.2}
\end{figure}
\newpage

\begin{figure}
\epsffile{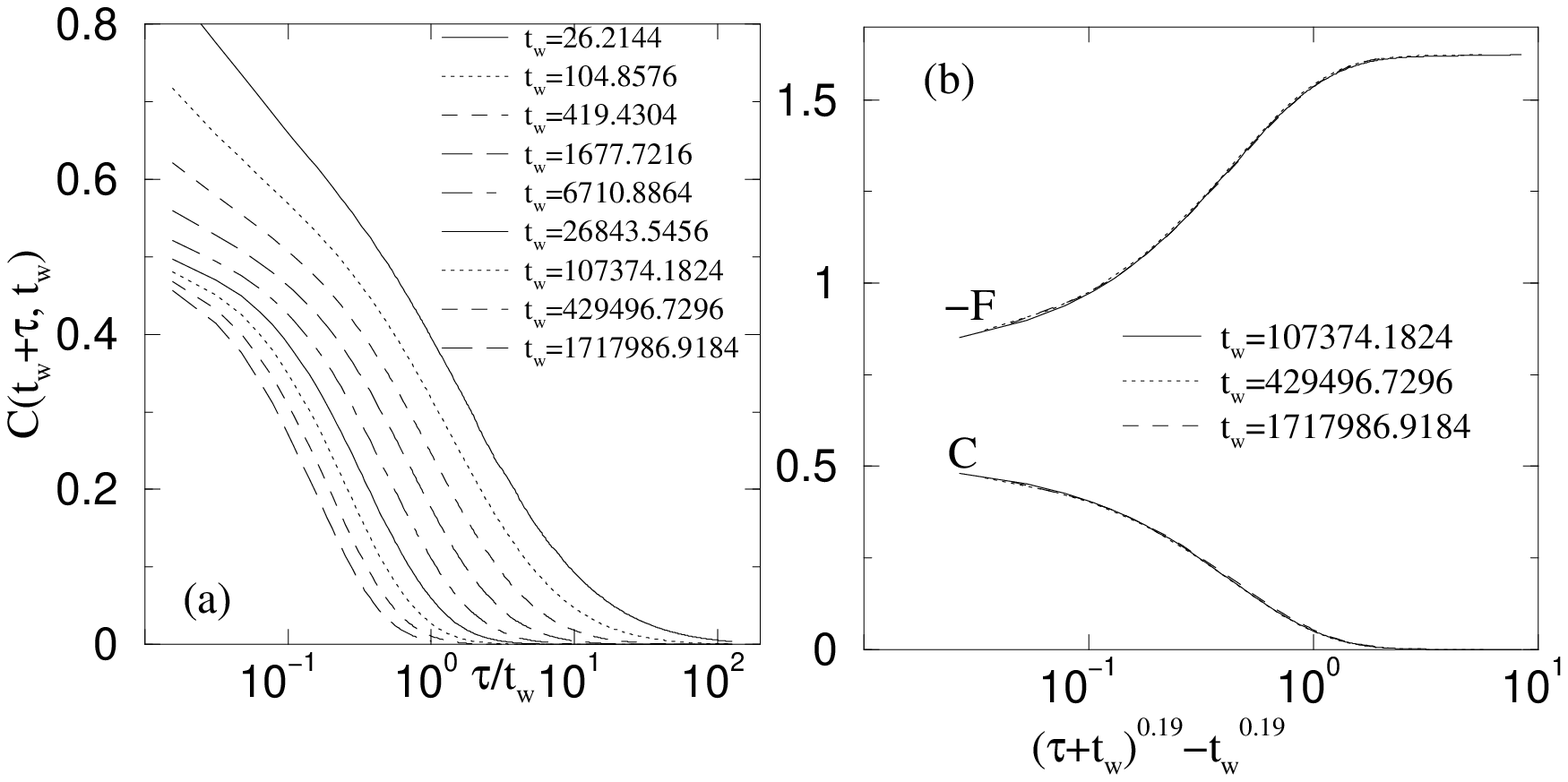}
\caption{(a)$C(t_w+\tau,t_w)$ versus rescaled time $\tau/t_w$ for 
$T=0.61$. (b) Scaling plot for $C(t_w+\tau,t_w)$ and $F(t_w+\tau,t_w)$
for $T=0.61$ using $h(t)=\exp[t^{1-\mu}/(1-\mu)]$ with $\mu=0.81$.}  
\label{f.3}
\end{figure}

\begin{thebibliography}{0}

\bibitem{kt}
   {Kirkpatrick T. R. and Thirumalai D.}, 
{Phys. Rev. B} {\bf 36} {(1987)} {5388}.

\bibitem{derr}
  {Derrida B.}, 
{Phys. Rev. Lett.} {\bf 45} {(1980)} {79}; 
   {Phys. Rev. B.} {\bf 24} {(1981)} {2613}.

\bibitem{gm}
  {Gross D. J. and M\'{e}zard M.}, {Nucl. Phys. B} {\bf 240} {(1984)} {431}. 


\bibitem{cs1}
   {Crisanti A. and Sommers H. J.}, {Z. Phys. B} {\bf 87} {(1992)} {341};
   {Crisanti A., Horner H.  and Sommers H. J.}, 
{Z. Phys. B} {\bf 92} {(1993)} {257}.

\bibitem{bckm1}
   {Bouchaud J. -P., Cugliandolo L. F.,  Kurchan J.  and M\'{e}zard
M.}, {Spin Glasses and Random Fields}, {A. P. Young}, {World
Scientific, Singapore} {(1998)}.

\bibitem{dh}
   {Deker U. and Haake F.}, {Phys. Rev. A} {\bf 11} {(1975)} {2043}.

\bibitem{h1} 
{Horner H.},  {Z. Phys. B} {\bf 87} {(1992)} {291}.


\bibitem{ck}
   {Cugliandolo L. F. and Kurchan J.}, {Phys. Rev. Lett.} {\bf 71}
{(1993)} {173}; 
    {J. Phys. A} {27} {(1994)} {5749};
    {Phil Mag. B} {\bf 71} {(1995)} {50}.

\bibitem{bckm2}
   {Bouchaud J. -P., Cugliandolo L. F.,  Kurchan J.  and M\'{e}zard
M.}, {Physica A} {\bf 226} {(1996)} {243}.




\bibitem{MSR}
   {Martin P. C., Siggia E. D. and  Rose H. A.}, 
{Phys. Rev. A} {\bf 8} {(1973)} {423}. 

\bibitem{cfik}
   {Chandra P., Feigelman M. V., Ioffe L. B., and Kagan D. M.}, 
{Phys. Rev. B} {\bf 56} {(1997)} {11553}.

\bibitem{fh}
   {Franz S. and Hertz J.}, {Phys. Rev. Lett.} {\bf 74} {(1995)} {2114}.

\bibitem{h2} 
{Kinzelbach H. and  Horner H.}, {J. Phys.I (France)} {\bf 3} {(1993)} {1392};
    {J. Phys.I (France)} {\bf 3} {(1993)} {1901}.


\bibitem{fm}
   {Franz S. and M\'{e}zard M.},  {Europhys. Lett.} {\bf 26} {(1994)} {209};
    {Physica A} {\bf 210} {(1994)} {48}.

\bibitem{cl}
   {Cugliandolo L. F.  and Le Doussal P.}, 
{Phys. Rev. E} {\bf 53} {(1996)} {1525}.

\bibitem{h3} 
{Horner H.}, {Z. Phys. B} {\bf 100} {(1996)} {243}.


\bibitem{gl}
   {Bengtzelius U., G\"{o}tze W. and Sj\"{o}lander A.}, 
{J. Phys. C} {\bf 17} {(1984)} {5915};
   {G\"{o}tze W.}, {Liquids, freezing and glass transition}, {Hansen
J. P., Levesque 
D. and Zinn-Justin J.}, {Les Houches} {(1989}.


\bibitem{leuth}
   {Leutheusser E.}, {Phys. Rev. A} {\bf 29} {(1984)} {2765}.

\bibitem{bbm}
   For the different situations, see
   {Barrat A., Burioni R.  and M\'{e}zard M.}, 
{J. Phys. A} {\bf 29} {(1996)} {L81}.

\bibitem{kpv}
   {Kurchan J., Parisi G. and Virasoro M. A.},  
{J. Phys. I} {\bf 3} {(1993)} {1819}.

\bibitem{cs2}
   {Crisanti A. and Sommers H. J.},  
{J. Phys. I} {\bf 5} {(1995)} {805}.

\bibitem{fp}
   {Franz S. and Parisi G.},  {J. Phys. I} {\bf 5} {(1995)} {1401}.

\bibitem{goetze}
   {G\"{o}tze W.},  {Z. Phys. B} {\bf 60} {(1985)} {195}.

\bibitem{fghl}
   {Fuchs M., G\"{o}tze, Hofacker H. and Latz. A}, 
{J. Phys. Condensed Matter} {\bf 3} {(1991)} {5047}.

\bibitem{parisi}
   {Parisi G.},  {Phys. Rev. Lett.} {\bf 79} {(1997)} {3660}.

\bibitem{kb}
   {Kob W. and Barrat J. -L. },  {Phys. Rev. Lett.} {\bf 78} {(1997)} {4581};
   {Barrat J. -L. and Kob W.},  {Europhys. Lett.} {\bf 46} {(1999)} {637};
   {Kob W. and Barrat J. -L. }
   Preprint cond-mat/9905248.

\bibitem{lapr}
   {Di Leonardo R., Angelani L., Parisi G. and Ruocco G},
   Preprint cond-mat/0001311.


\bibitem{rf}
   {Rieger H. and Franz S.},  {J. Stat. Phys.} {\bf 79} {(1995)} {749};
   {Marinari E., Parisi G., Ricci-Tersenghi F. and Ruiz-Lorenzo
J. J.},  {J. Phys. A} {\bf 31} {(1998)} {2611}.

\bibitem{zkh}
   {Zippold W., K\"{u}hn R. and Horner H.}, Preprint cond-mat/9904329.
   
\bibitem{vhob}
   {Vincent E.,  Hammann J., Ocio M. and Bouchaud J. -P.},
{Proceedings of the  Sitges Conference on Glassy Systems}, {E. Rubi},
{Springer, Berlin} {(1996)}.

\bibitem{latz00}
   {Latz A.}, J. Phys. Condens. Matt. (in print); preprint, {cond-mat/9911025}.
   
\end{thebibliography}
\end{document}